\documentclass[letterpaper, 10 pt, conference]{ieeeconf}
\IEEEoverridecommandlockouts                             
\usepackage{amsmath,amssymb}
\usepackage{graphicx,booktabs,array,cite}
\usepackage[dvipsnames]{xcolor}
\usepackage{tikz}
\usetikzlibrary{arrows.meta,positioning,calc,fit,shapes.misc,shapes.symbols,decorations.markings}
\usepackage{algorithm}
\usepackage[noend]{algpseudocode}
\algrenewcommand\algorithmicrequire{\textbf{Input}}
\algrenewcommand\algorithmicensure{\textbf{Output}}
\usepackage[hidelinks]{hyperref}

\makeatletter
\newcommand{\hd}[1]{\par\textbf{#1}\par\nobreak\@afterheading}
\makeatother
\newcommand{\Co}{\operatorname{Co}}
\newcommand{\tr}{\operatorname{tr}}
\newcommand{\re}{\Re}
\newcommand{\im}{\Im}

\newcommand{\adj}{\operatorname{adj}}
\newcommand{\spec}{\operatorname{spec}}
\newcommand{\num}{\operatorname{num}}
\newcommand{\C}{\mathbb{C}}
\newcommand{\R}{\mathbb{R}}
\newcommand{\N}{\mathbb{N}}
\newcommand{\RH}{\mathcal{RH}_\infty}
\newcommand{\cS}{\mathcal{S}}
\newcommand{\cC}{\mathcal{C}}
\newcommand{\cE}{\mathcal{E}}

\newcommand{\cX}{\mathcal{X}}

\newcommand{\cD}{\mathcal{D}}
\newcommand{\cR}{\mathcal{R}}
\providecommand{\mathdefault}[1]{#1}

\newtheorem{theorem}{Theorem}
\newtheorem{lemma}{Lemma}

\newtheorem{definition}{Definition}

\newtheorem{remark}{Remark}
\newtheorem{assumption}{Assumption}

\newcommand\mc[1]{{\color{black} #1}}

\begin{document}

\input{examplefig.tex}

\title{Decentralized stability analysis for networks of 2D one-port elements}

\author{Yi~An, Michael~Cantoni, and Farhad~Farokhi%
  \thanks{The first author is supported by an Australian Government
    Research Training Program Scholarship, and the Rowden White
    Scholarship.}%
  \thanks{The authors are all with the Department of
    Electrical and Electronic Engineering, The University of
    Melbourne, Parkville, VIC 3010,
    Australia. Email:~\texttt{an.y@student.unimelb.edu.au; \{cantoni;ffarokhi\}@unimelb.edu.au}}}


\maketitle

\begin{abstract}
  Networks of two-dimensional (2D) one-ports are relevant in the study
  of balanced AC power grids and two-axis vibration isolation, for
  example. The stability of a class of structured network
  interconnections is considered in this paper. Decentralized
  conditions on the elemental impedances of star networks are devised
  to ensure system stability when connected to coupling networks with
  given graph structured admittance. The results are enabled by
  positive operator-valued measure (POVM) \mc{based representations} of
  $2\times2$ matrices, borrowed from quantum information theory. These
  underlie the 2D generalization of existing results that are
  applicable to networks of only scalar (1D) flow and drop variables.
\end{abstract}


\section{Introduction}
\label{sec:intro}

The stability of interconnected dynamic networks is addressed in this
paper. A circuit theoretic perspective is adopted (e.g.,
see~\cite{Willems10}), with particular focus on networks of linear
time-invariant two-dimensional (2D) one-port elements. Such elements
involve a 2D flow through and 2D effort drop across two terminals. In
the frequency domain, the respective one-port impedance and admittance
are $2\times 2$ flow-to-drop and drop-to-flow transfer matrices.

A network is composed by connecting one-port element terminals at
junctions, which constrains the corresponding flow and drop variables
in accordance with Kirchhoff's laws. In particular, at every junction
the terminal effort variables become equal and the net flow becomes
zero. Networks of multi-dimensional one-ports are relevant in the
study of various engineered systems, including AC power
grids~\cite{Harnefors07,Gu21}, and vibration isolation
systems~\cite{Gardonio97}, for example.

\begin{figure}[!t]
\centering
\begin{tikzpicture}[>={Latex[length=1.8mm]}, font=\small, line width=0.5pt,
  blk/.style={draw, minimum width=12mm, minimum height=8mm, align=center}]
\node[blk] (Z) at (0, 0.7) {$-Z$};
\node[blk] (Y) at (0,-0.7) {$Y$};
\draw[->] (Z.east) -- (1.7,0.7) -- (1.7,-0.7) node[pos=0.5, right]{$e$} -- (Y.east);
\draw[->] (Y.west) -- (-1.7,-0.7) -- (-1.7,0.7) node[pos=0.5, left]{$f$} -- (Z.west);
\end{tikzpicture}
\caption{Feedback interconnection of the block graph Laplacian
  structured coupling network admittance
  $Y:=(B\otimes I_2)(\bigoplus_{\ell\in[q]} \gamma_\ell F) (B^*\otimes
  I_2)$ and star network impedance $Z:=\bigoplus_{k\in[p]} Z_k$. The
  return ratio is $L:=ZY$.}
\label{fig:loop}
\end{figure}
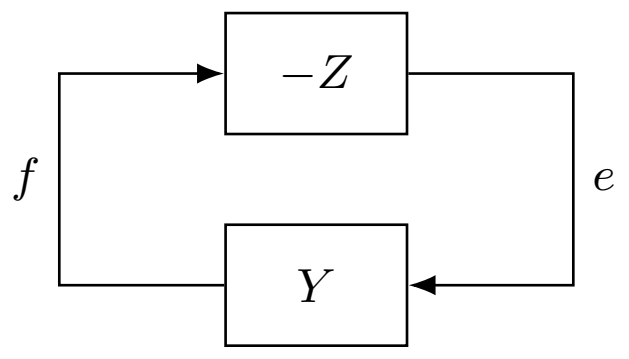

The specific system class considered involves the interconnection of
star networks and coupling networks with given graph structure, as
shown in Figure~\ref{fig:net}. This graph structure corresponds to an
internal network of 2D one-port elements, assumed to share a dynamic
factor.  For example, the star network could comprise linearized $dq$
co-ordinate models of inverter-based interfaces and other machines
across the buses of a balanced AC power grid, and the coupling network
the transmission lines with common resistance/reactance ratio. When
connected, the block-diagonal impedance of the star network is in
feedback with the block graph Laplacian structured admittance of the
coupling network, as shown in Figure~\ref{fig:loop}. Stability
verification is therefore amenable to Nyquist analysis and other
robust control tools~\cite{ZDG96,MR97,Vin01}.

In this paper, a Nyquist approach is adopted. Decentralized stability
conditions on the elemental $2\times 2$ impedances of the star network
are obtained in accordance with the graph structure of the coupling
network admittance. An analogous result for networks of only scalar
(1D) one-ports could be obtained from the convex-hull spectral
covering results in~\cite{LV06,Lestas12}. The generalization to 2D
one-ports is the main contribution of the developments presented
below. This is enabled by positive operator-valued measure (POVM)
\mc{based representations} of $2 \times 2$ matrices, borrowed from
quantum information theory~\cite{NC10}, and a corresponding geometric
interpretation that involves affine transformation of tetrahedral
encapsulations of the Bloch sphere~\cite{Renes04}.

\begin{figure}[!t]
\centering
{\begin{tikzpicture}[scale=0.75, font=\small, line width=0.5pt,
  >={Straight Barb[length=1.5mm, width=2.6mm]},
  term/.style={draw, fill=white, circle, inner sep=0pt, minimum size=4.2pt},
  dot/.style={fill, circle, inner sep=0pt, minimum size=3.2pt},
  flowmark/.style={postaction={decorate}, decoration={markings, mark=at position 0.62 with {\arrow{>}}}},
  inflow/.style={postaction={decorate}, decoration={markings, mark=at position 0.6 with {\arrow{>}}}}]
\def\ya{2.7} \def\yb{1.6} \def\yc{-0.8} \def\yd{-1.9}
\def\xs{2.1}   
\def\xt{2.9}   
\def\xb{4.2}   
\def\xw{5.52}  
\def\xn{5.8}   
\node[anchor=south, font=\footnotesize] at (0.9,3.4) {star network};
\coordinate (ref) at (0,0.4);
\draw[flowmark] (\xs,\ya) -- node[pos=0.36, anchor=south east, inner sep=1.5pt] {$-f_1$} (ref);
\draw[flowmark] (\xs,\yb) -- node[pos=0.24, anchor=north, inner sep=2.5pt] {$-f_2$} (ref);
\draw[flowmark] (\xs,\yc) -- node[pos=0.2, anchor=south, inner sep=4pt] {$-f_{p-1}$} (ref);
\draw[flowmark] (\xs,\yd) -- node[pos=0.36, anchor=north east, inner sep=1.5pt] {$-f_p$} (ref);
\node[dot, minimum size=3.6pt] at (ref) {};
\node[left=2pt, align=center, font=\footnotesize] at (ref) {ref.\\node $0$};
\foreach \y/\lab/\side in {\ya/1/above, \yb/2/above, \yc/p-1/below, \yd/p/below}{
  \draw (\xs,\y) -- (\xt,\y);
  \node[dot] at (\xs,\y) {};
  \node[term] at (\xt,\y) {};
  \node[\side=1.5pt] at (\xt,\y) {$e_{\lab}$};
}
\node at (\xt,0.45) {$\vdots$};
\draw[rounded corners=8pt] (\xw,-2.6) -- (\xw,3.4) -- (8.3,3.4) -- (8.3,-2.6) -- cycle;
\node[anchor=south, font=\footnotesize] at (6.9,3.4) {coupling network};
\foreach \y/\lab/\side/\nm in {\ya/1/above/a, \yb/2/above/b, \yc/p-1/below/c, \yd/p/below/d}{
  \node[term] (t\nm) at (\xb,\y) {};
  \node[\side=1.5pt] at (\xb,\y) {$e_{\lab}$};
  \draw[->] (t\nm) -- node[above, pos=0.45, inner sep=1.5pt] {$f_{\lab}$} (\xw,\y);
}
\node at (\xb,0.45) {$\vdots$};
\begin{scope}[shift={(6.9,0.4)}, line width=0.6pt]
  \draw (-0.6,1.0) -- (0.6,-1.0);
  \draw (-0.6,-1.0) -- (0.6,1.0);
  \draw (-0.85,1.0) -- (-0.6,1.0);  \draw (0.6,1.0) -- (0.85,1.0);
  \draw (-0.85,-1.0) -- (-0.6,-1.0); \draw (0.6,-1.0) -- (0.85,-1.0);
\end{scope}
\foreach \y in {\ya, \yb, \yc, \yd}{
  \draw[densely dashed, gray] (\xt+0.15,\y) -- (\xb-0.15,\y);
}
\end{tikzpicture}}
\caption{Interconnection of a star network comprising $p$ heterogeneous
  one-port elements (left) and a $p$-terminal coupling network
  (right). The reference node is the datum for all 2D effort variables
  $e_k$. When connected, the star network impedance and coupling
  network admittance both impose relationships between the 2D effort
  variables and 2D flow variables $f_k$. }
\label{fig:net}
\end{figure}
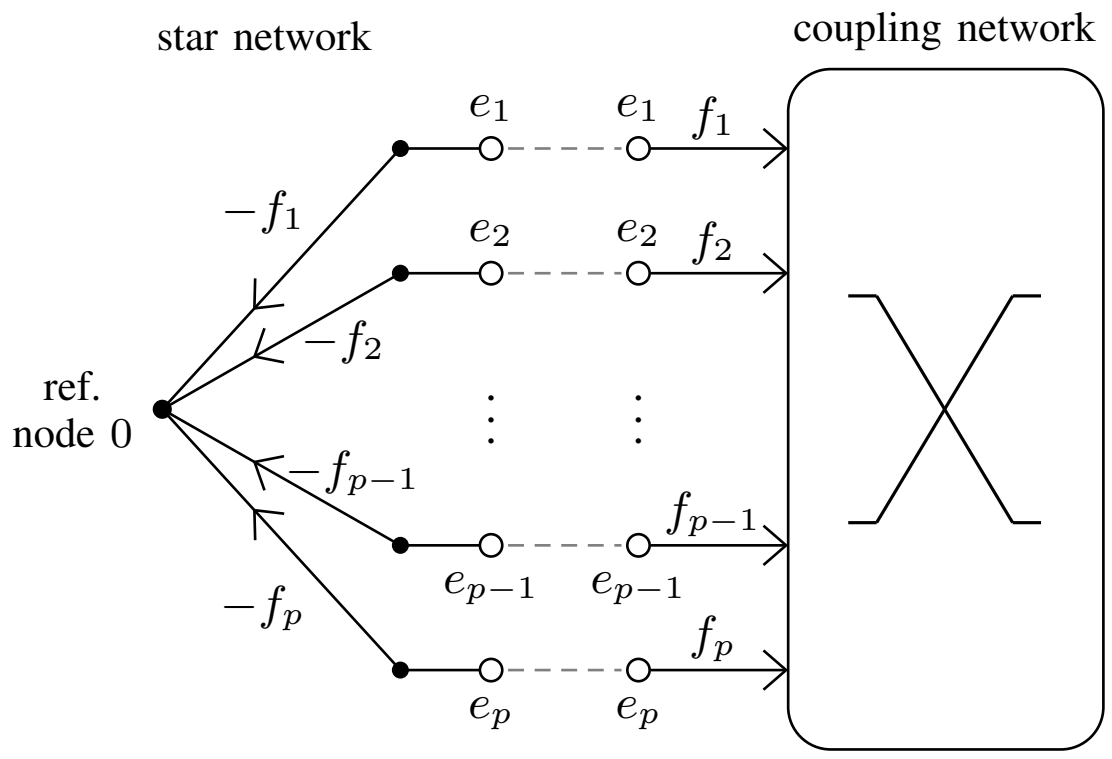

Decentralized stability conditions for feedback loops with block
network structure appear in various forms; e.g., integral quadratic
constraints~\cite{JK10,KR14,PV17}, mixed small gain/phase and
passivity~\cite{Huang24,HHHDL26,WCGPM26}, and scaled relative
graphs~\cite{BAD25}. The work in~\cite{PV12} and~\cite{HL26} is most
closely related to the subsequent developments. \mc{Both devise
  decentralized conditions via Nyquist criteria. In~\cite{PV12}, a
  generalized field-of-values is used to cover the spectrum of the
  structured return ratio. Optimizing the choice of underlying
  inner-product to mitigate conservativeness is understood only when
  the sub-systems are scalar. Here, the POVM-based representations
  used for the $2\times 2$ block elements are optimized by
  re-orienting a related tetrahedral encapsulation of the Bloch
  sphere, which does not scale the numerical range. Further, the
  elemental POVMs lead to a {\em scalar} decomposition of
  the return ratio.
  A corresponding {\em block} decomposition in~\cite{HL26} is by
  contrast tailored to the graph structure alone.} On the other
hand,~\cite{PV12} and~\cite{HL26} accommodate full heterogeneity in
the dynamics of coupling network elements. Such a broadening of scope
and the incorporation of field-of-values scaling within the POVM
approach developed below are part of ongoing work.

Basic notation and preliminaries are given next. The stability
analysis problem is formulated in Section~\ref{sec:problem}. A
POVM-based decentralized spectral covering result and decentralized
stability conditions are presented in Section~\ref{sec:main}. The
POVMs best suited to these conditions are characterized in
Section~\ref{sec:povm}. Aspects of the results are illustrated
numerically in Section~\ref{sec:example}, within the context of a
simple AC power network. To conclude, future work is discussed in
Section~\ref{sec:conclusion}.

\section{Preliminaries}
\label{sec:prelim}
\subsection{Notation}
The natural, real, and complex numbers are $\N$, $\R$, and $\C$.
Ordered $[n]\!:=\!\{1,\dots,n\}\subset\mathbb{N}$,
\mc{$[\theta_1,\theta_2]\!:=\!\{\theta:\theta_1\!\leq\! \theta\!\leq\! \theta_2\}$
  and ${]}\theta_1,\theta_2{[}\!:=\!\{\theta:\theta_1 \!<\!
  \theta \!<\! \theta_2\}$}
are real intervals, $j:=\sqrt{-1}$, and $\Re(\zeta)$ is the real part
of $\zeta\in\C$.
The (column) vector $\mathbf{1}_n:=(1,\ldots,1)\in\C^n$, and
$I_n\in\C^{n\times n}$ is the identity matrix. For
$X\in\C^{n \times m}$, conjugate transpose is denoted by
$X^*\in\C^{m \times n}$, maximum singular value by
$\sigma_{\max}(X)\geq 0$, and element-wise modulus by
\mc{$|X|\in\R^{n\times m}$.} Given $\{X_i\}_{i\in[n]}$,
block-diagonal
$\bigoplus_{i\in[n]}X_i\!:=\!X_{1}\oplus\cdots\oplus X_{n}$, and
$X_1\otimes X_2$ denotes Kronecker product. For
$X \in \C^{n \times n}$, the spectrum is
$\spec(X):=\{\lambda\in\C:\det(\lambda I_n - X)=0\}$, where
$\det(\cdot)$ denotes determinant, the numerical range
(field-of-values) is $\num(X):=\{x^*Xx:x^*x=1\}\subset\C$, and
$\tr(\cdot)$ denotes trace. If $X^*X=XX^*=I_n$, then $X$ is called
unitary. A unit $x\in\C^n$ satisfies $\|x\|:=\sqrt{x^*x}=1$. The set
$\cR$ consists of rational transfer functions, and $\RH$ the proper
ones free of poles with non-negative real part.

\subsection{Graphs}
\label{sec:graphs}
For \mc{connected} graphs with $p\in\N\backslash \{1\}$ nodes, $q\in\N$
edges, $B\in\{-1,0,1\}^{p\times q}$ denotes the incidence matrix \mc{for
some edge orientation, and $B_{k,\ell}$ the entry in row $k\in[p]$, column
$\ell\in[q]$. Further, $\cE_k:=\{\ell:B_{k,\ell}\neq 0\}$, with non-zero
cardinality $|\cE_k|$ by connectedness.} Given
$\gamma_\ell > 0$ and $\Gamma:=\bigoplus_{\ell\in[q]}\gamma_\ell$, the
weighted graph Laplacian is $B\Gamma B^*$.

\subsection{Convex hulls}
\label{sec:shull}
The set $\cX\subset\C$ is convex if
$\theta \alpha + (1-\theta)\beta\in\cX$ for all $\alpha,\beta\in\cX$,
$0\leq \theta \leq 1$. For
$\cX\!\subset\!\C$, the convex hull
$\Co(\cX)\!:=\! \{\sum_{i\in[n]}\!\theta_i \alpha_i: n\!\in\!\N,
\alpha_i\!\in\!\cX,\, \theta_i\!\geq\! 0,\,
\sum_{i\in[n]}\!\theta_i\!=\!1\}$ is the smallest convex set
containing $\cX$~\cite[App.~B]{HornJohnson}.
\subsection{POVM-based representations of $2\times 2$ matrices}
\label{sec:povmprelim}

\mc{The set of rank-1 matrices $\{\mu_mu_mu_m^*:m\in[4]\}$,} with unit
$u_m\in\C^2$ and $\mu_m>0$, is a POVM~\cite[Sec.~2.2.6]{NC10} if
\begin{align}
\sum_{m\in[4]}\mu_mu_mu_m^*=I_2.
\label{eq:complete}
\end{align}
\mc{The tuple of coefficients
$(\varphi_{m})_{m\in[4]}\in\C\times\C\times\C\times\C$} is a
corresponding \mc{POVM-based} representation of $\Phi\in\C^{2\times 2}$ if
\begin{align}
  \Phi=\sum_{m\in[4]}\varphi_{m}\,\mu_{m}u_{m}u_{m}^* = U^*\Big(\bigoplus_{m\in[4]}\varphi_{m}\Big)U,
  \label{eq:compression}
\end{align}
where
$U:=(\bigoplus_{m\in[4]}\sqrt{\mu_m}u_m^*)(\mathbf{1}_4\otimes I_2)$.
\mc{This representation always uniquely exists if
  $\sum_{m\in[4]}\alpha_m u_mu_m^*=0$ implies $\alpha_m=0$,
  $m\in[4]$. In this case, the POVM is minimally informationally
  complete~\cite{Renes04}.}
\begin{lemma}
\label{lem:cover}
If \mc{$(\varphi_m)_{m\in[4]}$} is a \mc{POVM-based} representation
of $\Phi\in\C^{2\times 2}$, then
$\num(\Phi)\subset\Co(\{\varphi_m:m\in[4]\})$.
\end{lemma}
\begin{proof}
  For any unit $x\in\C^2$, the non-negative scalars
  $\mu_{m}|u_{m}^*x|^2$ sum to $x^*x=1$
  by~\eqref{eq:complete}, making
  $x^*\Phi x=\sum_{m\in[4]}\varphi_{m}\,\mu_{m}|u_{m}^*x|^2$ a convex
  combination of the $\varphi_m$.
\end{proof}

\section{Problem formulation}
\label{sec:problem}
\subsection{Interconnected network stability}

As introduced in Section~\ref{sec:intro}, the problem of interest is
stability analysis for interconnected networks of 2D one-ports. In
particular, the connection of a star network with $p$ elemental 2D
one-ports and a coupling network with given graph structure, as shown
in Figure~\ref{fig:net}. The graph structure of the coupling network
reflects the internal interconnection of 2D one-port elements, assumed
to share a dynamic factor, as formalized in
Assumption~\ref{ass:struct} subsequently.

\mc{When connected as shown in Figure~\ref{fig:net},} the star network and
coupling network both constrain the 2D flow variables
$f_1,\ldots,f_p$, and 2D effort variables $e_1,\ldots, e_p$. \mc{The
elements of the} star network impose \mc{the impedance} relationships
$e_k = -Z_k f_k$ with $Z_k\in\cR^{2\times 2}$, $k\in[p]$. The
coupling network imposes \mc{the admittance} relationships
\begin{align}
   \begin{bmatrix}
        f_1\\\vdots\\f_p
    \end{bmatrix} = Y \begin{bmatrix}
        e_1\\\vdots\\ e_p
    \end{bmatrix}. 
\end{align}
The star network impedance $Z:=\bigoplus_{k\in[p]} Z_k$ is therefore
in feedback with the coupling admittance $Y$, as shown in
Figure~\ref{fig:loop}, where $e=(\oplus_{k\in[p]}\mc{e_k})\mathbf{1}_p$ and
$f=(\oplus_{k\in[p]}\mc{f_k})\mathbf{1}_p$.
\begin{assumption}
\label{ass:struct}
The following are assumed throughout:
\begin{enumerate}
\item The graph structured coupling network admittance is 
\begin{equation}
Y=(B\otimes I_2)\big(\textstyle \bigoplus_{\ell\in[q]}Y_\ell\big)(B^*\otimes I_2).
\label{eq:Y}
\end{equation}
where $B$ is \mc{a corresponding} incidence matrix. 
\item \label{ass:common}
Each $Y_\ell=\gamma_\ell F$, with $F\in\RH^{2\times2}$ and $\gamma_\ell > 0$. 
\item \label{ass:stable} The return ratio $L:=ZY\in\RH^{2p\times 2p}$
  \mc{under preclusion of unstable pole-zero cancellation.}
\end{enumerate}
\end{assumption}
\begin{definition}
\label{def:stab}
The interconnection of the star network and the coupling network is
stable if $(I_{\mc{2p}}+L)^{-1}\in\RH^{\mc{2p\times 2p}}$.
\end{definition}

\subsection{A centralized stability condition}
\begin{lemma}
\label{lem:homotopy}
If
\begin{equation}
(\forall \omega\in\R\cup\{\infty\})( \forall\, \tau\in[0,1])~
-1\notin\spec\big(\tau L(j\omega)\big),
\label{eq:homotopy}
\end{equation}
then $(I_{\mc{2p}}+L)^{-1}\in\RH^{\mc{2p\times 2p}}$. 
\end{lemma}

A proof is given in Appendix~\ref{app:homotopy}, based on a homotopy
argument similar to one found in~\cite[p.~17]{Vin01}.
\begin{remark}
\label{rem:bound}
Condition \eqref{eq:homotopy} is centralized as it involves every star
element and the whole coupling network.
For each $\omega\in\mathbb{R}\cup\{\infty\}$, suppose
$\cC_\omega\subset\C$ is convex, with $0\in\cC_\omega$,
$-1\notin\cC_\omega$, and $\spec(L(j\omega))\subset\cC_\omega$. Then
$\spec(\tau
L(j\omega))=\tau\,\spec(L(j\omega))\subset\cC_\omega\not\ni-1$ for all
$\tau\in[0,1]$, \mc{$\omega\in\R\cup\{\infty\}$, and hence,
  $(I_{2p}+L)^{-1}\in\RH^{\mc{2p\times 2p}}$ by
  Lemma~\ref{lem:homotopy}. The problem of interest is to find
  decentralized conditions on the star network elements that imply
  the existence of such sets $\cC_\omega$. The following lemma is a 
  step in this direction.}
\end{remark}

\begin{lemma}
\label{lem:sym}
Let $\Phi:=\bigoplus_{k\in[p]}\Phi_k$ with $\Phi_k:=Z_k F$, and 
\[
  M:=(\Gamma^{1/2}\otimes I_2)(B^*\otimes I_2)\,
  \Phi\,(B\otimes I_2)(\Gamma^{1/2}\otimes I_2).
\]
Then $\spec(L(j\omega))\cup\{0\}=\spec(M(j\omega))\cup\{0\}$ \mc{for all
$\omega$.}
\end{lemma}

A proof is given in Appendix~\ref{app:sym}.  Unlike
$L(j\omega)\in\C^{2p\times 2p}$, which lives over the coupling graph
nodes, the permuted composition $M(j\omega)\in\C^{2q\times 2q}$
lives over the edges.


\section{Main results}
\label{sec:main}
\subsection{Decentralized spectral covering}
\label{sec:cover}
\begin{theorem}
  \label{thm:dspc}
  \mc{Given $\omega\in\R\cup\{\infty\}$,}  let
  $(\varphi_{k,m}(j\omega))_{m\in[4]}$ be \mc{a POVM-based representation}
  of $\Phi_k(j\omega)$, $k\in[p]$. Then
\begin{align*}
  \spec(M(j\omega))&\subset\num(M(j\omega))\\
                   &\subset\Co\Big(
                     \!\bigcup_{k\in[p]}\bigcup_{m\in[4]}\!
                     \{0,\frac{\max_{\ell\in\cE_k}\gamma_\ell}{\chi_k^2}\,
                     \varphi_{k,m}(j\omega)\}\!\Big)
\end{align*}
where $\chi_k:=\min\{1/|\cE_k|,1/2\}$.
\end{theorem}

A proof is given in Appendix~\ref{app:main}. Theorem~\ref{thm:dspc}
extends~\cite[Lemma~2]{LV06} from scalar elements to $2\times2$ block
elements. The cover of $\spec(M(j\omega))$ involves 
sets that each depend on only $\Phi_k(j\omega)$
and the edge weights $\{\gamma_\ell:\ell\in\cE_k\}$.

\subsection{Decentralized stability conditions}
\label{sec:cert}
The following lemma is an instance of the separating hyperplane
theorem~\cite[Sec.~2.5.1]{BV04}, given $\C\sim\R^2$. The
separating line through $-1$, at angle $-\theta$ from \mc{vertical (i.e.,
$\theta$ clockwise),} is
$\{\zeta:\re(\exp(j\theta)\zeta)=-\cos\theta\}\subset\C$.  This is
used to decentralize Theorem~\ref{thm:dspc} in line
with~\cite[Rem.~2]{LV06}.
\begin{lemma}
\label{lem:line}
Given closed and convex $\cC\subset\C$ with $0\in\cC$,
$-1\!\notin\!\cC \Leftrightarrow
(\exists \theta \!\in\,
]\tfrac{-\pi}{2},\!\tfrac{\pi}{2}[\,)
( \forall
\zeta\!\in\!\cC)~\re(\exp(j\theta)\zeta)\!>\!-\cos\theta$.
\end{lemma}

\begin{theorem}
\label{thm:cert}
For each $\omega\in\R\cup\{\infty\}$, let
$(\varphi_{k,m}(j\omega))_{m\in[4]}$ be \mc{a POVM-based} representation of
$\Phi_k(j\omega)$, \mc{$k\in[p]$}. Further, suppose $\mc{(\forall \omega\in\R\cup\{\infty\})\,(\exists \theta \!\in\,
  ]\tfrac{-\pi}{2},\!\tfrac{\pi}{2}[\,)}$
\begin{align}
  \label{eq:local}
  \begin{split}
&\mc{(\forall (k,m)\in[p]\times[4])}
  \\
& \qquad\quad \frac{\max_{\ell\in\cE_k}\gamma_\ell}{\chi_k^2}\,
  \re\big(\exp(j\theta)\varphi_{k,m}(j\omega)\big)>-\cos\theta.
  \end{split}
\end{align}
Then the interconnected system is stable.
\end{theorem}
\begin{proof}
  Given
  $\omega\in\R\cup\{\infty\}$, let \mc{$\cC_\omega:=\Co(\{0\}\cup
    \{(\max_{\ell\in\cE_k}\gamma_\ell/\chi_k^2)\varphi_{k,m}(j\omega):
    (k,m)\in[p]\times[4]\})$. With any
    $\theta\in{]}\frac{-\pi}{2},\frac{\pi}{2}{[}$ such
    that~\eqref{eq:local} holds, it follows by the
    $\R$-linearity of $\Re(\cdot)$ that every
    $\zeta\in\cC_\omega$ satisfies}
  $\re(\exp(j\theta)\zeta)>-\cos\theta$, and thus,
  $-1\notin\cC_\omega$ by Lemma~\ref{lem:line}. Further, by
  Lemma~\ref{lem:sym} and Theorem~\ref{thm:dspc},
  $\spec(L(j\omega))\subset\spec(M(j\omega))\cup\{0\}\subset\cC_\omega$.
  \mc{The corresponding collection of sets
    $\{\cC_\omega:\omega\in\R\cup\{\infty\}\}$ complies with the
    sufficient stability condition given in Remark~\ref{rem:bound},
    whereby} $(I_{2p}+L)^{-1}\in\RH^{\mc{2p\times 2p}}$.
\end{proof}

\section{Choosing the POVMs}
\label{sec:povm}
\mc{At every frequency, and for each star network element, the choice of
POVM is free. It is shown below how to optimize the choice for given
candidate angle $\theta$ in~\eqref{eq:local}.}

\subsection{The Bloch sphere and a reference POVM}
\label{sec:ball}
Every unit $x\in\C^2$ corresponds to a vector $v(x)$ on the unit
(Bloch) sphere in $\R^3$~\cite[Sec.~1.2]{NC10}. With the Pauli
matrices $H_0:=I_2$,
$H_1:=\left[\begin{smallmatrix}0&1\\1&0\end{smallmatrix}\right]$,
$H_2:=\left[\begin{smallmatrix}0&-j\\j&0\end{smallmatrix}\right]$,
$H_3:=\left[\begin{smallmatrix}1&0\\0&-1\end{smallmatrix}\right]$, and
$\nu_i(x):=x^*H_i x$ for $i\in\{0\}\cup[3]$, this Bloch vector is
$v(x):=(\nu_1(x),\nu_2(x),\nu_3(x))\in\R^3$ and $\nu_0(x)=1$. Further,
\[
x^*\Phi_kx=T_k(v(x)):=\phi_{k,0}+\textstyle\sum_{i\in[3]}\phi_{k,i}\nu_i(x),
\]
where $\phi_{k,i}:=\frac{1}{2}\tr(\Phi_k H_i)\mc{\in\C}$. The image of the
sphere under affine $T_k$ is $\num(\Phi_k)$, as elaborated in
Appendix~\ref{app:sic}.

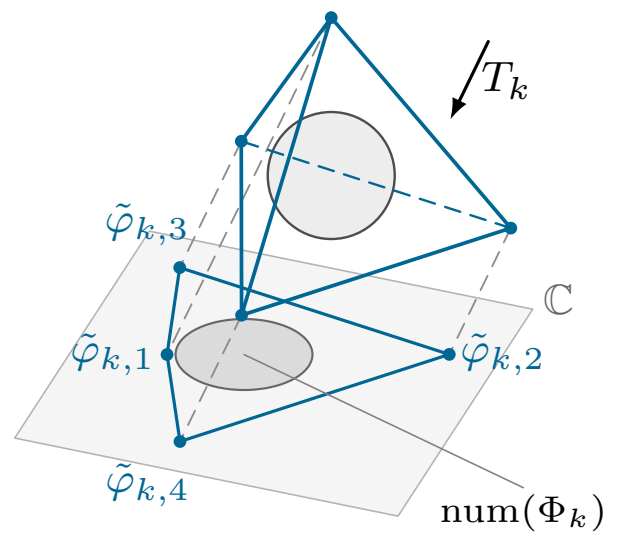
\begin{figure}[!t]
  \centering
  \vspace*{10pt}
\begin{tikzpicture}[x=0.42cm, y=0.42cm, font=\footnotesize, line cap=round, line join=round, >={Latex[length=1.4mm]}]
\fill[black!4] (-4.985,-4.135) -- (1.052,-5.364) -- (3.174,-2.105) -- (-2.864,-0.876) -- cycle;
\draw[black!30, line width=0.3pt] (-4.985,-4.135) -- (1.052,-5.364) -- (3.174,-2.105) -- (-2.864,-0.876) -- cycle;
\fill[black!14] (-0.980,-2.298) -- (-1.069,-2.282) -- (-1.160,-2.271) -- (-1.253,-2.263) -- (-1.347,-2.260) -- (-1.441,-2.261) -- (-1.534,-2.266) -- (-1.627,-2.275) -- (-1.717,-2.289) -- (-1.805,-2.306) -- (-1.889,-2.328) -- (-1.970,-2.353) -- (-2.046,-2.382) -- (-2.117,-2.414) -- (-2.182,-2.449) -- (-2.241,-2.487) -- (-2.294,-2.527) -- (-2.339,-2.570) -- (-2.378,-2.614) -- (-2.408,-2.661) -- (-2.431,-2.708) -- (-2.445,-2.756) -- (-2.452,-2.805) -- (-2.450,-2.854) -- (-2.440,-2.902) -- (-2.422,-2.950) -- (-2.396,-2.997) -- (-2.362,-3.042) -- (-2.321,-3.086) -- (-2.272,-3.128) -- (-2.217,-3.167) -- (-2.155,-3.204) -- (-2.087,-3.238) -- (-2.014,-3.269) -- (-1.936,-3.296) -- (-1.854,-3.319) -- (-1.768,-3.339) -- (-1.679,-3.355) -- (-1.587,-3.367) -- (-1.495,-3.374) -- (-1.401,-3.378) -- (-1.307,-3.377) -- (-1.213,-3.372) -- (-1.121,-3.362) -- (-1.030,-3.349) -- (-0.942,-3.331) -- (-0.858,-3.310) -- (-0.777,-3.285) -- (-0.701,-3.256) -- (-0.630,-3.224) -- (-0.565,-3.189) -- (-0.506,-3.151) -- (-0.453,-3.110) -- (-0.408,-3.068) -- (-0.370,-3.023) -- (-0.339,-2.977) -- (-0.317,-2.930) -- (-0.302,-2.881) -- (-0.295,-2.833) -- (-0.297,-2.784) -- (-0.307,-2.735) -- (-0.325,-2.688) -- (-0.351,-2.641) -- (-0.385,-2.595) -- (-0.426,-2.551) -- (-0.475,-2.510) -- (-0.530,-2.470) -- (-0.592,-2.433) -- (-0.660,-2.399) -- (-0.733,-2.369) -- (-0.811,-2.342) -- (-0.894,-2.318) -- cycle;
\draw[black!60, line width=0.4pt] (-0.980,-2.298) -- (-1.069,-2.282) -- (-1.160,-2.271) -- (-1.253,-2.263) -- (-1.347,-2.260) -- (-1.441,-2.261) -- (-1.534,-2.266) -- (-1.627,-2.275) -- (-1.717,-2.289) -- (-1.805,-2.306) -- (-1.889,-2.328) -- (-1.970,-2.353) -- (-2.046,-2.382) -- (-2.117,-2.414) -- (-2.182,-2.449) -- (-2.241,-2.487) -- (-2.294,-2.527) -- (-2.339,-2.570) -- (-2.378,-2.614) -- (-2.408,-2.661) -- (-2.431,-2.708) -- (-2.445,-2.756) -- (-2.452,-2.805) -- (-2.450,-2.854) -- (-2.440,-2.902) -- (-2.422,-2.950) -- (-2.396,-2.997) -- (-2.362,-3.042) -- (-2.321,-3.086) -- (-2.272,-3.128) -- (-2.217,-3.167) -- (-2.155,-3.204) -- (-2.087,-3.238) -- (-2.014,-3.269) -- (-1.936,-3.296) -- (-1.854,-3.319) -- (-1.768,-3.339) -- (-1.679,-3.355) -- (-1.587,-3.367) -- (-1.495,-3.374) -- (-1.401,-3.378) -- (-1.307,-3.377) -- (-1.213,-3.372) -- (-1.121,-3.362) -- (-1.030,-3.349) -- (-0.942,-3.331) -- (-0.858,-3.310) -- (-0.777,-3.285) -- (-0.701,-3.256) -- (-0.630,-3.224) -- (-0.565,-3.189) -- (-0.506,-3.151) -- (-0.453,-3.110) -- (-0.408,-3.068) -- (-0.370,-3.023) -- (-0.339,-2.977) -- (-0.317,-2.930) -- (-0.302,-2.881) -- (-0.295,-2.833) -- (-0.297,-2.784) -- (-0.307,-2.735) -- (-0.325,-2.688) -- (-0.351,-2.641) -- (-0.385,-2.595) -- (-0.426,-2.551) -- (-0.475,-2.510) -- (-0.530,-2.470) -- (-0.592,-2.433) -- (-0.660,-2.399) -- (-0.733,-2.369) -- (-0.811,-2.342) -- (-0.894,-2.318) -- cycle;
\draw[MidnightBlue, line width=0.6pt] (-2.586,-2.819) -- (-2.382,-4.189) -- (1.859,-2.817) -- (-2.386,-1.450) -- cycle;
\draw[black!45, densely dashed, line width=0.35pt] (0.000,2.487) -- (-2.586,-2.819);
\draw[black!45, densely dashed, line width=0.35pt] (2.828,-0.828) -- (1.859,-2.817);
\draw[black!45, densely dashed, line width=0.35pt] (-1.417,0.540) -- (-2.386,-1.450);
\draw[black!45, densely dashed, line width=0.35pt] (-1.412,-2.200) -- (-2.382,-4.189);
\fill[black!7] (0.000,0.000) circle (1);
\draw[black!70, line width=0.45pt] (0.000,0.000) circle (1);
\draw[MidnightBlue, densely dashed, line width=0.45pt] (2.828,-0.828) -- (-1.417,0.540);
\draw[MidnightBlue, line width=0.7pt] (0.000,2.487) -- (2.828,-0.828);
\draw[MidnightBlue, line width=0.7pt] (0.000,2.487) -- (-1.417,0.540);
\draw[MidnightBlue, line width=0.7pt] (0.000,2.487) -- (-1.412,-2.200);
\draw[MidnightBlue, line width=0.7pt] (2.828,-0.828) -- (-1.412,-2.200);
\draw[MidnightBlue, line width=0.7pt] (-1.417,0.540) -- (-1.412,-2.200);
\fill[MidnightBlue] (0.000,2.487) circle (1.2pt);
\fill[MidnightBlue] (2.828,-0.828) circle (1.2pt);
\fill[MidnightBlue] (-1.417,0.540) circle (1.2pt);
\fill[MidnightBlue] (-1.412,-2.200) circle (1.2pt);
\fill[MidnightBlue] (-2.586,-2.819) circle (1.2pt);
\fill[MidnightBlue] (1.859,-2.817) circle (1.2pt);
\fill[MidnightBlue] (-2.386,-1.450) circle (1.2pt);
\fill[MidnightBlue] (-2.382,-4.189) circle (1.2pt);
\node[MidnightBlue, inner sep=0pt] at (-3.436,-2.819) {$\tilde{\varphi}_{k,1}$};
\node[MidnightBlue, inner sep=0pt] at (2.709,-2.817) {$\tilde{\varphi}_{k,2}$};
\node[MidnightBlue, inner sep=0pt] at (-2.892,-0.766) {$\tilde{\varphi}_{k,3}$};
\node[MidnightBlue, inner sep=0pt] at (-2.885,-4.874) {$\tilde{\varphi}_{k,4}$};
\draw[->, line width=0.6pt] (2.473,2.110) -- (1.874,0.880) node[right, inner sep=2pt, pos=0.5] {$T_k$};
\node[anchor=south east, black!60, inner sep=2pt] at (4.004,-2.366) {$\C$};
\draw[black!50, line width=0.3pt] (-1.374,-2.819) -- (3.027,-4.917) node[below, inner sep=1pt, black, font=\scriptsize] {$\num(\Phi_k)$};
\end{tikzpicture}
\caption{Tetrahedron around the Bloch sphere. Under
  $T_k:\R^3\rightarrow\C$, the vertex shadows are
  $\{\tilde{\varphi}_{k,m}:m\in[4]\}$, and the sphere shadow is
  $\num(\Phi_k)$.}
\label{fig:proj}
\end{figure}

Define $\mu_m:=\frac{1}{2}$, $m\in[4]$, and the following units in $\C^2$:
\begin{align}
  \begin{split}
    \tilde{u}_1&:=\big(1,\, 0 \big), \qquad \quad ~\,
                 \tilde{u}_3:=
                 \big(\tfrac{1}{\sqrt{3}},\,
                 \tfrac{\sqrt{2}}{\sqrt{3}}\exp(j\tfrac{2\pi}{3})\big);
                 \\
    \tilde{u}_2&:=\big(\tfrac{1}{\sqrt{3}},\,
                 \tfrac{\sqrt{2}}{\sqrt{3}}\big), \qquad 
                 \tilde{u}_4:=\big(\tfrac{1}{\sqrt{3}},\,
                 \tfrac{\sqrt{2}}{\sqrt{3}}\exp(j\tfrac{4\pi}{3})\big).
  \end{split}
      \label{eq:referenceU}
\end{align}
The vectors $\tilde{v}_m:=v(\tilde{u}_m)\in\mathbb{R}^3$ yield a
regular tetrahedron $\Co(\{3\tilde{v}_m:m\in[4]\})$ that encapsulates
the Bloch sphere. The tuple $(\tilde{\varphi}_{k,m})_{m\in[4]}$ with
$\tilde{\varphi}_{k,m}:=T_k(3\tilde{v}_m)\in\C$, $m\in[4]$, is a
representation of $\Phi_k$ \mc{with respect to} the reference POVM
$\{\mu_m\tilde{u}_m\tilde{u}_m^*:m\in[4]\}$, \mc{which is
  minimally informationally complete.} See Fig.~\ref{fig:proj}, and
Appendix~\ref{app:sic} for details.

\subsection{Turning the encapsulation to align with candidate
  \texorpdfstring{$\theta$}{theta}}
\label{sec:turn}
Let $b_{k,\theta}:=(\re(\exp(j\theta)\phi_{k,i}))_{i\in[3]}$ and
$(\mc{\beta_{k,i}})_{i\in[3]}:=\|b_{k,\theta}\|^{-1}b_{k,\theta}\in\R^3$. If
$b_{k,\theta}=0$, then let $V_{k,\theta}:=I_2$, else if $\beta_{k,3}=-1$,
then let $V_{k,\theta}:=H_1$, otherwise let
\begin{align} \label{eq:Vrot} V_{k,\theta}\!:=\!\!
  \tfrac{1}{\sqrt{2(1+\beta_{k,3})}}
  \big((1\!+\!\beta_{k,3})I_2\!+\!j(\beta_{k,2}H_1\!-\!\beta_{k,1}H_2)\big).
\end{align}
Note that $V_{k,\theta}$ is unitary, and define
$u_{k,m}(\theta):=V_{k,\theta}\,\tilde{u}_m$, $m\in[4]$. With
$v_{k,m}(\theta):=v(u_{k,m}(\theta))$, the turned representation of
$\Phi_k$ has coefficients $\varphi_{k,m}:=T_k(3v_{k,m}(\theta))$.
\begin{theorem}
  \label{thm:aligned} For $m\in\{2,3,4\}$,
  $\re(\exp(j\theta)\varphi_{k,m})=
  \min\{\re(\exp(j\theta)\zeta):\zeta\in\num(\Phi_k)\}
  \leq\re(\exp(j\theta)\varphi_{k,1})$.
\end{theorem}

\begin{figure}[!t]
  \centering
  \vspace*{10pt}
\begin{tikzpicture}[font=\footnotesize, line cap=round, line join=round, >={Latex[length=1.4mm]}]
\begin{scope}[x=0.5cm, y=0.5cm]
\fill[black!7] (0,0) circle (1);
\draw[black!45, densely dotted, line width=0.35pt] (0.999,0.008) -- (0.995,0.024) -- (0.986,0.040) -- (0.973,0.055) -- (0.956,0.071) -- (0.935,0.086) -- (0.910,0.100) -- (0.881,0.115) -- (0.848,0.128) -- (0.812,0.141) -- (0.772,0.154) -- (0.728,0.166) -- (0.682,0.177) -- (0.633,0.187) -- (0.581,0.197) -- (0.526,0.206) -- (0.469,0.214) -- (0.411,0.221) -- (0.350,0.227) -- (0.288,0.232) -- (0.225,0.236) -- (0.161,0.239) -- (0.096,0.241) -- (0.031,0.242) -- (-0.035,0.242) -- (-0.100,0.241) -- (-0.165,0.239) -- (-0.229,0.235) -- (-0.292,0.231) -- (-0.354,0.226) -- (-0.415,0.220) -- (-0.473,0.213) -- (-0.530,0.205) -- (-0.584,0.196) -- (-0.636,0.187) -- (-0.685,0.176) -- (-0.731,0.165) -- (-0.774,0.153) -- (-0.814,0.140) -- (-0.850,0.127) -- (-0.883,0.114) -- (-0.912,0.099) -- (-0.937,0.085) -- (-0.958,0.070) -- (-0.974,0.054) -- (-0.987,0.039) -- (-0.995,0.023) -- (-1.000,0.007);
\draw[black!55, line width=0.35pt] (-0.999,-0.008) -- (-0.995,-0.024) -- (-0.986,-0.040) -- (-0.973,-0.055) -- (-0.956,-0.071) -- (-0.935,-0.086) -- (-0.910,-0.100) -- (-0.881,-0.115) -- (-0.848,-0.128) -- (-0.812,-0.141) -- (-0.772,-0.154) -- (-0.728,-0.166) -- (-0.682,-0.177) -- (-0.633,-0.187) -- (-0.581,-0.197) -- (-0.526,-0.206) -- (-0.469,-0.214) -- (-0.411,-0.221) -- (-0.350,-0.227) -- (-0.288,-0.232) -- (-0.225,-0.236) -- (-0.161,-0.239) -- (-0.096,-0.241) -- (-0.031,-0.242) -- (0.035,-0.242) -- (0.100,-0.241) -- (0.165,-0.239) -- (0.229,-0.235) -- (0.292,-0.231) -- (0.354,-0.226) -- (0.415,-0.220) -- (0.473,-0.213) -- (0.530,-0.205) -- (0.584,-0.196) -- (0.636,-0.187) -- (0.685,-0.176) -- (0.731,-0.165) -- (0.774,-0.153) -- (0.814,-0.140) -- (0.850,-0.127) -- (0.883,-0.114) -- (0.883,-0.114) -- (0.912,-0.099) -- (0.937,-0.085) -- (0.958,-0.070) -- (0.974,-0.054) -- (0.987,-0.039) -- (0.995,-0.023) -- (1.000,-0.007);
\draw[black!70, line width=0.45pt] (0,0) circle (1);
\draw[MidnightBlue, densely dashed, line width=0.45pt] (0.000,2.911) -- (-0.491,-0.296);
\draw[MidnightBlue, densely dashed, line width=0.45pt] (2.658,-1.204) -- (-0.491,-0.296);
\draw[MidnightBlue, densely dashed, line width=0.45pt] (-0.491,-0.296) -- (-2.167,-1.410);
\draw[MidnightBlue, line width=0.7pt] (0.000,2.911) -- (2.658,-1.204);
\draw[MidnightBlue, line width=0.7pt] (0.000,2.911) -- (-2.167,-1.410);
\draw[MidnightBlue, line width=0.7pt] (2.658,-1.204) -- (-2.167,-1.410);
\draw[->, line width=0.6pt] (0,0) -- (0,1) node[right, inner sep=2.5pt, pos=0.5] {};
\node[anchor=south, inner sep=1.5pt] at (0.000,2.911) {$3(\beta_{k,1},\beta_{k,2},\beta_{k,3})$};
\end{scope}
\begin{scope}[xshift=3.685cm, x=1.3cm, y=1.3cm]
\draw[black!35, line width=0.3pt, ->] (-1.450,0) -- (1.360,0) node[below left, inner sep=1pt, black!60] {$\re$};
\draw[black!35, line width=0.3pt, ->] (0,-0.818) -- (0,1.497) node[below left, inner sep=1pt, black!60] {$\im$};
\fill[black!7] (0.378,0.116) -- (0.376,0.145) -- (0.372,0.175) -- (0.365,0.207) -- (0.354,0.241) -- (0.339,0.277) -- (0.320,0.314) -- (0.296,0.352) -- (0.267,0.390) -- (0.233,0.427) -- (0.195,0.462) -- (0.153,0.494) -- (0.109,0.522) -- (0.064,0.545) -- (0.018,0.564) -- (-0.027,0.579) -- (-0.070,0.588) -- (-0.111,0.594) -- (-0.149,0.595) -- (-0.184,0.594) -- (-0.216,0.589) -- (-0.245,0.583) -- (-0.272,0.575) -- (-0.297,0.564) -- (-0.319,0.553) -- (-0.339,0.540) -- (-0.357,0.526) -- (-0.373,0.511) -- (-0.388,0.495) -- (-0.401,0.478) -- (-0.413,0.460) -- (-0.423,0.440) -- (-0.431,0.420) -- (-0.438,0.398) -- (-0.444,0.374) -- (-0.447,0.349) -- (-0.448,0.323) -- (-0.447,0.294) -- (-0.443,0.264) -- (-0.435,0.231) -- (-0.425,0.197) -- (-0.410,0.162) -- (-0.391,0.124) -- (-0.366,0.087) -- (-0.337,0.049) -- (-0.304,0.012) -- (-0.266,-0.023) -- (-0.224,-0.055) -- (-0.180,-0.083) -- (-0.134,-0.107) -- (-0.088,-0.126) -- (-0.044,-0.140) -- (-0.000,-0.149) -- (0.040,-0.155) -- (0.078,-0.156) -- (0.113,-0.155) -- (0.146,-0.151) -- (0.175,-0.144) -- (0.202,-0.136) -- (0.226,-0.126) -- (0.248,-0.114) -- (0.268,-0.101) -- (0.287,-0.088) -- (0.303,-0.073) -- (0.318,-0.056) -- (0.331,-0.039) -- (0.342,-0.021) -- (0.352,-0.002) -- (0.361,0.019) -- (0.368,0.041) -- (0.373,0.064) -- (0.376,0.089) -- cycle;
\draw[black!70, line width=0.45pt] (0.378,0.116) -- (0.376,0.145) -- (0.372,0.175) -- (0.365,0.207) -- (0.354,0.241) -- (0.339,0.277) -- (0.320,0.314) -- (0.296,0.352) -- (0.267,0.390) -- (0.233,0.427) -- (0.195,0.462) -- (0.153,0.494) -- (0.109,0.522) -- (0.064,0.545) -- (0.018,0.564) -- (-0.027,0.579) -- (-0.070,0.588) -- (-0.111,0.594) -- (-0.149,0.595) -- (-0.184,0.594) -- (-0.216,0.589) -- (-0.245,0.583) -- (-0.272,0.575) -- (-0.297,0.564) -- (-0.319,0.553) -- (-0.339,0.540) -- (-0.357,0.526) -- (-0.373,0.511) -- (-0.388,0.495) -- (-0.401,0.478) -- (-0.413,0.460) -- (-0.423,0.440) -- (-0.431,0.420) -- (-0.438,0.398) -- (-0.444,0.374) -- (-0.447,0.349) -- (-0.448,0.323) -- (-0.447,0.294) -- (-0.443,0.264) -- (-0.435,0.231) -- (-0.425,0.197) -- (-0.410,0.162) -- (-0.391,0.124) -- (-0.366,0.087) -- (-0.337,0.049) -- (-0.304,0.012) -- (-0.266,-0.023) -- (-0.224,-0.055) -- (-0.180,-0.083) -- (-0.134,-0.107) -- (-0.088,-0.126) -- (-0.044,-0.140) -- (-0.000,-0.149) -- (0.040,-0.155) -- (0.078,-0.156) -- (0.113,-0.155) -- (0.146,-0.151) -- (0.175,-0.144) -- (0.202,-0.136) -- (0.226,-0.126) -- (0.248,-0.114) -- (0.268,-0.101) -- (0.287,-0.088) -- (0.303,-0.073) -- (0.318,-0.056) -- (0.331,-0.039) -- (0.342,-0.021) -- (0.352,-0.002) -- (0.361,0.019) -- (0.368,0.041) -- (0.373,0.064) -- (0.376,0.089) -- cycle;
\draw[BrickRed, densely dashed, line width=0.55pt] (-0.522,1.209) -- (-0.588,0.210) -- (-0.208,-0.613) -- (1.178,0.072) -- cycle;
\draw[MidnightBlue, line width=0.7pt] (1.097,-0.502) -- (-0.005,1.166) -- (-0.822,-0.249) -- cycle;
\draw[black, line width=0.6pt] (-1.310,-0.537) -- (-0.260,1.282);
\draw[black!60, densely dotted, line width=0.4pt] (-1,0) -- (-1,0.78);
\draw[black!70, line width=0.4pt, -{Latex[length=1.1mm, width=1mm]}] (-1,0.6) arc (90:60:0.6);
\node[inner sep=0.5pt] at (-0.808,0.715) {$\theta$};
\draw[BrickRed, line width=0.45pt, fill=white] (-0.522,1.209) circle (1.2pt);
\draw[BrickRed, line width=0.45pt, fill=white] (-0.588,0.210) circle (1.2pt);
\draw[BrickRed, line width=0.45pt, fill=white] (-0.208,-0.613) circle (1.2pt);
\draw[BrickRed, line width=0.45pt, fill=white] (1.178,0.072) circle (1.2pt);
\fill[MidnightBlue] (1.097,-0.502) circle (1.3pt);
\fill[MidnightBlue] (-0.411,0.463) circle (1.3pt);
\fill[MidnightBlue] (-0.822,-0.249) circle (1.3pt);
\fill[MidnightBlue] (-0.005,1.166) circle (1.3pt);
\draw[black] (-1,0) node[cross out, draw, inner sep=1.3pt, line width=0.5pt] {};
\node[anchor=south east, inner sep=1.5pt] at (-1.03,0.02) {$-1$};
\draw[black!50, line width=0.3pt] (0.200,-0.060) -- (0.520,-0.640) node[below, inner sep=1pt, black, font=\scriptsize] {$\num(\Phi_k)$};
\node[anchor=west, inner sep=2pt, MidnightBlue, font=\scriptsize] at (1.097,-0.502) {$\varphi_{k,1}$};
\node[anchor=west, inner sep=2pt, MidnightBlue, font=\scriptsize] at (-0.411,0.463) {$\varphi_{k,2}$};
\node[anchor=north, inner sep=2pt, MidnightBlue, font=\scriptsize] at (-0.880,-0.249) {$\varphi_{k,3}$};
\node[anchor=west, inner sep=2.5pt, MidnightBlue, font=\scriptsize] at (-0.005,1.166) {$\varphi_{k,4}$};
\end{scope}
\end{tikzpicture}
\caption{Turning the tetrahedron, the transformation under $T_k$ is a
  triangle (solid) with one edge on the supporting line of
  $\num(\Phi_k)$ at angle $\theta$. For the reference POVM (dashed)
  the separating line is crossed.}
\label{fig:turn}
\end{figure}
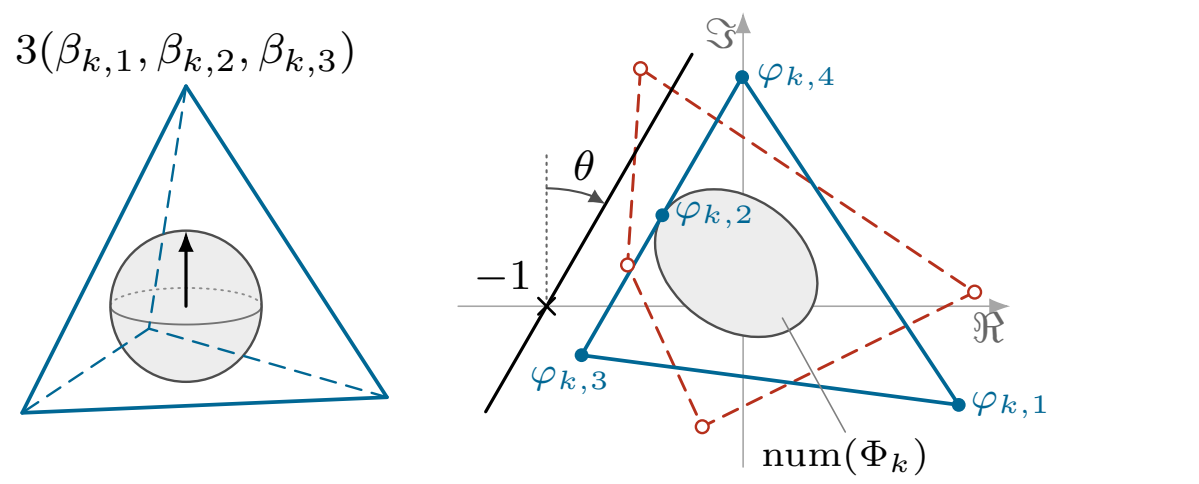

A proof is given in Appendix~\ref{app:aligned}. Note that the convex
hull $\Co(\{\varphi_{k,m}:m\in[4]\})$ is a triangle with one side on
the support of $\num(\Phi_k)$ in direction $\theta$, as shown in
Figure~\ref{fig:turn}. Therefore, the least conservative application of
Theorem~\ref{thm:cert} involves search over
$\theta\in{]}\frac{-\pi}{2},\frac{\pi}{2}{[}$ to
verify~\eqref{eq:local} with correspondingly turned representations of
the $\Phi_k(j\omega)$, $k\in[p]$, frequency-by-frequency.

\section{Example: AC power grid, uniform \texorpdfstring{$R/X$}{R/X} lines}
\label{sec:app}\label{sec:example}

\begin{figure}[!t]
\centering
\resizebox{0.9\columnwidth}{!}{\begin{tikzpicture}[line width=0.6pt, font=\small,
  >={Straight Barb[length=1.5mm, width=2.6mm]},
  bus/.style={line width=1.6pt},
  dev/.style={draw, fill=white, minimum width=14mm, minimum height=10mm, align=center, inner sep=1.5pt},
  upflow/.style={postaction={decorate}, decoration={markings, mark=at position 0.8 with {\arrow{>}}}}]
\draw[bus] (0,0) -- (2,0);
\draw[bus] (4.2,0) -- (6.2,0);
\draw[bus] (8.4,0) -- (10.4,0);
\draw (1.6,0) -- (1.6,0.6) -- node[above] {$\gamma_2 F$} (4.6,0.6) -- (4.6,0);
\draw (5.8,0) -- (5.8,0.6) -- node[above] {$\gamma_1 F$} (8.8,0.6) -- (8.8,0);
\draw[upflow] (1,-1.2) -- node[right, pos=0.35, inner sep=2pt] {$f_2$} (1,0);
\node[anchor=north east, inner sep=2pt] at (0.8,-0.05) {$e_2$};
\node[dev, anchor=north] at (1,-1.2) {$Z_2$\\[-1pt]GFM};
\draw[upflow] (5.2,-1.2) -- node[right, pos=0.35, inner sep=2pt] {$f_1$} (5.2,0);
\node[anchor=north east, inner sep=2pt] at (5.0,-0.05) {$e_1$};
\node[dev, anchor=north] at (5.2,-1.2) {$Z_1$\\[-1pt]GFL};
\draw[upflow] (9.4,-1.08) -- node[right, pos=0.35, inner sep=2pt] {$f_3$} (9.4,0);
\node[anchor=north east, inner sep=2pt] at (9.2,-0.05) {$e_3$};
\node[cloud, draw, fill=white, cloud puffs=11, aspect=1.7, inner sep=0.5pt, align=center]
  at (9.4,-1.65) {$Z_3=0$\\[-1pt]stiff grid};
\end{tikzpicture}}
\caption{The test system: GFL inverter $Z_1$, GFM inverter $Z_2$ and
  stiff grid $Z_3=0$, connected by two lines with admittances
  $\gamma_1F$ and $\gamma_2F$.}
\label{fig:test}
\end{figure}
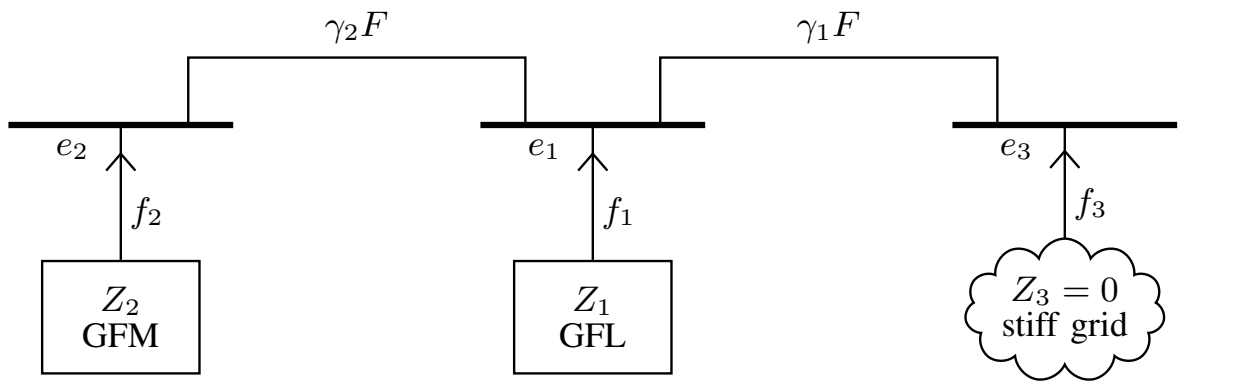

Figure~\ref{fig:test} shows a grid-following (GFL) inverter at
bus~$1$, a grid-forming (GFM) inverter at bus~$2$, and a stiff grid
connection at bus~$3$. Each line admittance is $\gamma_\ell F(s)$ with
$\gamma_\ell:=1/X_\ell$, $F(s):=((s/\omega_0+\varrho)I_2-jH_2)^{-1}$,
and common $\varrho:=R_\ell/X_\ell$ (e.g., see~\cite{Huang24}), given
resistance $R_\ell$ and reactance $X_\ell$ for $\ell\in\{1,2\}$,
\mc{and $dq$-frame angular speed $\omega_0/(2\pi)=50$\,Hz. The
  inverters have operating-point dependent impedances $Z_1$ and $Z_2$,
  and the stiff-bus impedance is $Z_3=0$.}

\begin{figure}[!t]
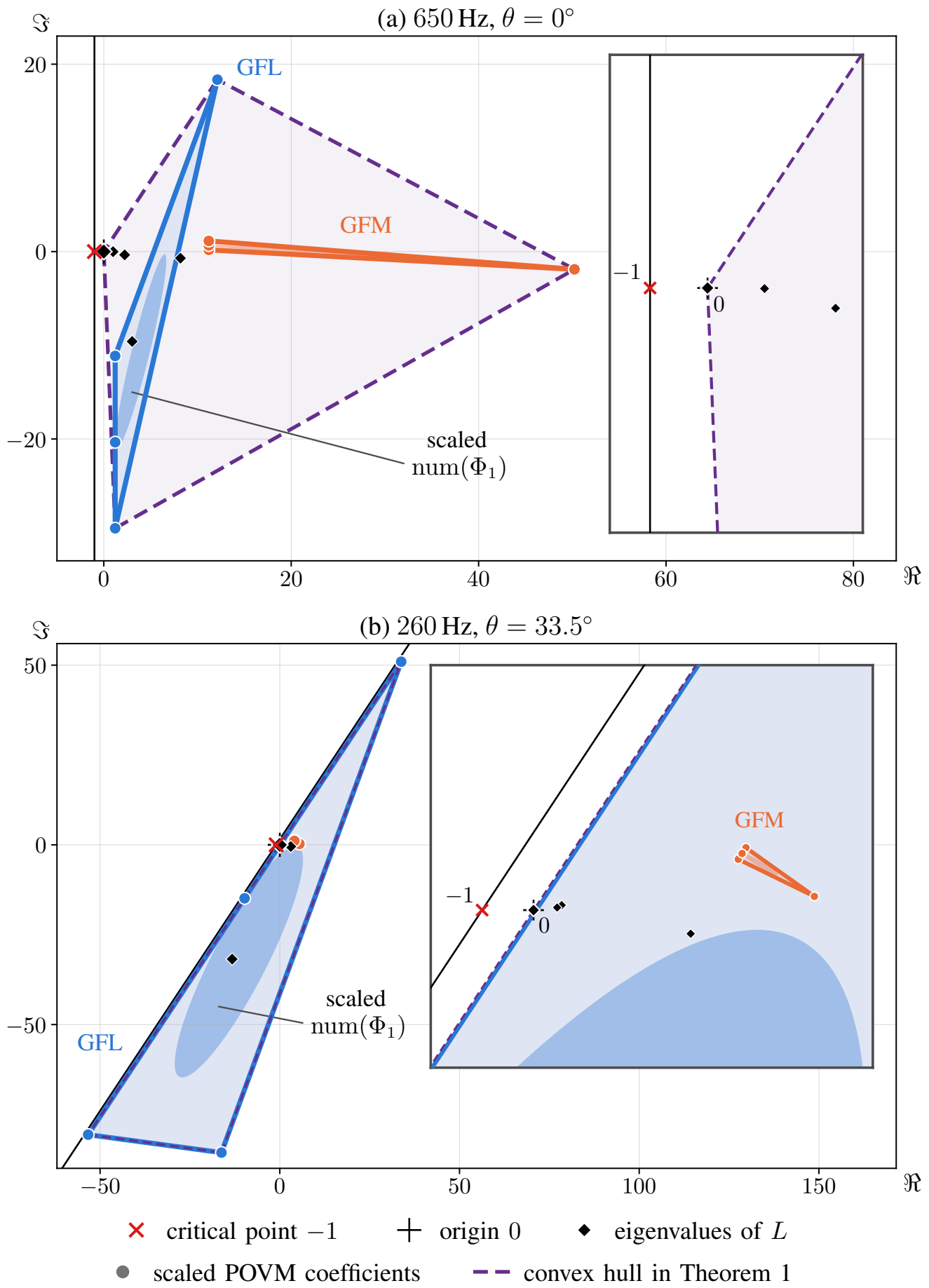

\centering
\resizebox{\columnwidth}{!}{\usebox\figexbox}
\caption{Objects associated with verifying~\eqref{eq:local} for
  different $\theta$ at two different frequencies, noting that the
  condition always holds for the stiff grid element $\Phi_3=Z_3F=0$:
  (i) Elemental convex hulls of POVM coefficients scaled by
  corresponding $\max_{\ell\in\cE_k}\gamma_\ell/\chi_k^2$, for GFL
  $\Phi_1=Z_1F$ (blue $k=1$), and GFM $\Phi_2=Z_2F$ (orange $k=2$);
  (ii) Convex hull in Theorem~\ref{thm:dspc} (purple), which contains
  $\spec(L(j\omega))$ (black diamonds).}
\label{fig:fixed}
\end{figure}

Two instances of the objects associated with~\eqref{eq:local} are
shown in Figure~\ref{fig:fixed} for two frequencies $\omega/(2\pi)$
\mc{and relevant line parameters and inverter impedances}: $650$\,Hz
with $\theta=0^\circ$; $260$\,Hz with $\theta=33.5^\circ$. The
triangle vertices for each inverter are the POVM coefficients
$\varphi_{k,m}$, turned to align with $\theta$, and the elliptical
region is the numerical range of $\Phi_k$, both scaled by
$\max_{\ell\in\cE_k}\gamma_\ell/\chi_k^2$. The line separates all from
$-1$ at both frequencies. So the convex hull in
Theorem~\ref{thm:dspc}, which contains \mc{all six (counting
  multiplicity) elements of} $\spec(L)$, excludes $-1$ by
Lemma~\ref{lem:line}. At $650$\,Hz, a \mc{passivity-like} vertical
line through $-1$ suffices. This is not the case at $260$\,Hz.

\section{Conclusion}
\label{sec:concl}\label{sec:conclusion}
Decentralized conditions on the elemental 2D one-ports of star
networks are obtained to ensure system stability when connected to
coupling networks with graph structure. This is enabled by utilizing
POVM-based representations of $2\times 2$ matrices, borrowed from
quantum information theory. POVM choice offers flexibility to tame
conservativeness. Next steps include accommodating: (i) fully
heterogeneous coupling dynamics; (ii) generalized field-of-values
scaling of star impedances; and (iii) unstable star impedances.

\appendices

\section{Proof of Lemma~\ref{lem:homotopy}}
\label{app:homotopy}
The result holds by an argument similar to~\cite[p.~17]{Vin01}, via
the principle of the argument \cite[Thm.~A.1.3]{CZ95}, which underpins
the Nyquist criterion~\cite[Thm.~A.1.14]{CZ95}.

Consider $\tau\mapsto\det(I_{2p}+\tau L)$ over $\tau\in[0,1]$. Each
$\det(I_{2p}+\tau L)$ is a scalar rational function \mc{with no poles
  on $j\R$, and a constant limit at $\infty$, because
  $L\in\RH^{2p\times2p}$. Further,}
$\omega\mapsto\det(I_{2p}+\tau L(j\omega))$ over
$\omega\in\R\cup\{\infty\}$ is a closed \mc{smooth} curve that
\mc{deforms continuously with} $\tau$. \mc{Let
  $\mathrm{wno}(\det(I_{2p}+\tau L))$ denote the number of times this
  curve winds around the origin in the clockwise direction as $\omega$
  increases along $\mathbb{R}$.}

The hypothesis~\eqref{eq:homotopy} implies
$\det\big(I_{2p}+\tau L(j\omega)\big) \neq 0$ for all
$\omega\in\mathbb{R}\cup\{\infty\}$, $\tau\in[0,1]$.  As such,
$\tau\mapsto\mathrm{wno}\det(I_{2p}+\tau L)$ must be constant on the
interval $[0,1]$ because for the winding number to change the
determinant must become zero at some frequency. This constant is
$\mathrm{wno}\det(I_{2p})=0=\mathrm{wno}\det(I_{2p}+L)$. By the
principle of the argument, $n_z-n_p=\mathrm{wno}\det(I_{2p}+L)=0$,
where $n_z$ and $n_p$ are respectively the number of zeros and poles
of $\det(I_{2p}+L)$ in $\{\zeta\in\C~:~\Re(\zeta)> 0\}$. \mc{Since
  $L\in\RH^{2p\times2p}$ implies $n_p=0$, it follows that $n_z=0$.}
Therefore,
$(I_{2p}+L)^{-1}= \frac{1}{\det(I_{2p}+L)} \adj(I_{2p}+L)
\in\RH^{2p\times2p}$.

\section{Proof of Lemma~\ref{lem:sym}}
\label{app:sym}
By the properties of Kronecker
products~\cite[Sec.~4.2]{HornJohnsonTopics}, and
Assumption~\ref{ass:struct},
$Y=(B\otimes I_2) (\Gamma \otimes F) (B^*\otimes I_2) = (B \Gamma B^*)
\otimes F = (I_p \otimes F)((B \Gamma B^*) \otimes I_2)$. As such,
$L=ZY=(\bigoplus_{k\in[p]} Z_k F)((B\Gamma B^*)\otimes I_2)=\Phi\,((B
\Gamma B^*)\otimes I_2)$. Since
$(B\Gamma B^*)\otimes I_2=(B\otimes I_2)(\Gamma^{1/2}\otimes
I_2)(\Gamma^{1/2}\otimes I_2)(B \otimes I_2)^*$, it follows that the
non-zero spectrum of
\[
  M(j\omega) = (\Gamma^{1/2}\otimes I_2)(B \otimes I_2)^*
  \Phi(j\omega) (B\otimes I_2)(\Gamma^{1/2}\otimes I_2)
\] 
coincides with the non-zero spectrum of $L(j\omega)$ because
$\spec(PQ)\cup\{0\}=\spec(QP)\cup\{0\}$ for all complex matrices
$P\in\C^{m\times n}$ and
$Q\in\C^{n\times m}$~\cite[p.65]{HornJohnson}.

\section{Proof of Theorem~\ref{thm:dspc}} 
\label{app:main}

Inclusion of the spectrum in the numerical range is well
known~\cite[p.10]{HornJohnsonTopics}. The second inclusion is a
consequence of a convex-hull cover of the latter, as established below
via the following scalar result from~\cite[Lem.~1]{LV06}, given
$\Phi_k=U_k^*(\bigoplus_{m\in[4]}\varphi_{k,m}) U_k$, where the $m$-th
row of $U_k$ is $\sqrt{\mu_{k,m}}\,u_{k,m}^*$ in line
with~\eqref{eq:compression}:
\begin{lemma}
\label{lem:LV}
\mc{For $\cD\!\subset\!\C$, let
  $\cS(\cD)\!:=\!\{\zeta^2:\zeta\!\in\!\Co(\{\delta:\delta^2\!\in\!\cD\})\}\!\ni\!
  0$ denote the S-hull, which is convex. Further, given
  $r\!\in\!\N\backslash\{1\}$, let $\delta_t\!\in\!\C$ for
  $t\!\in\![r]$.} Then $\big\{
(\sum_{t\in[r]} \delta_t\alpha_t^*)
(\sum_{t\in[r]} \delta_t\alpha_t)\big/
\mc{\eta^2} :
 \mc{\eta=
\sum_{t\in[r]} |\alpha_t| 
\neq 0},\,
\alpha_t\in\C \big\} =
\cS\big(\{\delta_t^2:t\in[r]\}\big)$.
\end{lemma}

Consider any $x=(\bigoplus_{\ell\in[q]} x_\ell)\mathbf{1}_q$ with
$x_\ell\in\C^2$ such that
$\|x\|^2=x^*x=\sum_{\ell\in[q]}x_\ell^*x_\ell=1$, and for $k\in[p]$, define
\[
  y_k:=\sum_{\ell\in[q]}\sqrt{\gamma_\ell}\, B_{k,\ell}U_kx_\ell \in
  \C^4.
\]
Then
$(\bigoplus_{k\in[p]}y_k)\mathbf{1}_p=
(\bigoplus_{k\in[p]}U_k)(B\otimes I_2)(\Gamma^{1/2}\otimes I_2)x$,
and
\[
  x^*Mx=\sum_{k\in[p]}y_k^*(\bigoplus_{m\in[4]}\varphi_{k,m})y_k,
\]
since $\Gamma^{1/2}$ has real entries. Given $\chi_k:=\min\{1/|\cE_k|,1/2\}$,
\[
  \alpha_{k,\ell,m}:=\chi_kB_{k,\ell}\sqrt{\mu_{k,m}}\,u_{k,m}^*x_\ell,
\]
$a_{k,\ell}:=(\bigoplus_{m\in[4]}|\sqrt{\mu_{k,m}}\,u_{k,m}^*x_\ell|)\mathbf{1}_4\in\R^4$,
and
\begin{align}
  \cC:=\Co\big(\bigcup_{k\in[p]}\bigcup_{m\in[4]}\frac{\varphi_{k,m}}{\chi_k^2}\,
  \cS(\{\gamma_\ell:\ell\in\cE_k\})\big),
\label{eq:CuShulls}
\end{align}
it follows that
\begin{align}
  x^*Mx&=\!\sum_{k\in[p]}\sum_{m\in[4]}\!\frac{\varphi_{k,m}}{\chi_k^2}
          \big(\sum_{\ell\in[q]}\!\!\sqrt{\gamma_\ell}\,\alpha_{k,\ell,m}^*\big)
         \!\big(\sum_{\ell\in[q]}\!\!\sqrt{\gamma_\ell}\,\alpha_{k,\ell,m}\big)
         \nonumber\\
       &\in\Big(\sum_{k\in[p]}\sum_{m\in[4]}
         \big(\sum_{\ell\in[q]}\!|\alpha_{k,\ell,m}|\big)^2\Big)\,\cC\nonumber\\
       &=\Big(\sum_{k\in[p]}\big\|\sum_{\ell\in[q]}\chi_k|B_{k,\ell}|\,
         a_{k,\ell}\big\|^2\Big)\,\cC\nonumber\\
       &\subset\Big(\sum_{k\in[p]}\Big(\sum_{\ell\in[q]}\chi_k|B_{k,\ell}|\,\|x_\ell\|\Big)^2\Big)\,\cC \subset \cC.
\label{eq:numbound}
\end{align}
In particular, the first inclusion holds by Lemma~\ref{lem:LV} if
$q>1$, and directly if $q=1$. The second inclusion holds by the
triangle inequality, and the positivity of $\chi_k$, since
$\|a_{k,\ell}\|^2=x_\ell^*(\sum_{m\in[4]}\mu_{k,m}u_{k,m}u_{k,m}^*)x_\ell
=\|x_\ell\|^2$
by the completeness of POVMs encoded in~\eqref{eq:complete}.
With $\rho_k$ and $\kappa_\ell$ as the $k$-th row and $\ell$-th column
sums of entry-wise non-negative $(\bigoplus_{k\in[p]}\chi_k)|B|$, and
$\tilde{x}:=(\bigoplus_{\ell\in[q]}\|x_\ell\|)\mathbf{1}_q\in\R^q$,
the final inclusion holds because
\begin{align*}
&\sum_{k\in[p]}\Big(\sum_{\ell\in[q]}\chi_k|B_{k,\ell}|\,\|x_\ell\|\Big)^2\\
&\quad = 
\Big(\big(\bigoplus_{k\in[p]}\chi_k\big)\,|B|\,\tilde{x} \Big)^*
\Big(\big(\bigoplus_{k\in[p]}\chi_k\big)\,|B|\,\tilde{x}\Big)\\
&\quad \le\sigma_{\max}\Big(\Big(\bigoplus_{k\in[p]}\chi_k\Big)|B|\Big)^2
  \|\tilde{x}\|^2
\le\max_{k\in[p],\,\ell\in[q]}\rho_k\kappa_\ell \leq 1.
\end{align*}
To see this, note that
$\|\tilde{x}\|^2=\sum_{\ell\in[q]}\|x_\ell\|^2=\|x\|^2=1$. Further,
the second inequality is Schur's maximum singular value
bound~\cite{Schur11}, with $\rho_k=\chi_k|\cE_k|\le1$, and
$\kappa_\ell=(\chi_i+\chi_{i^\prime})\le 1$ for the edge indexed by
$\ell$ between nodes $i$ and $i^\prime$, because
$\chi_k \le 1/|\cE_k|$ and $\chi_k\le\frac{1}{2}$ for every
$k\in[p]$. Finally, the real edge weights $\gamma_\ell$ are positive,
and so the S-hulls in~\eqref{eq:CuShulls} are the intervals
$\cS(\{\gamma_\ell:\ell\in\cE_k\})=[0,\max_{\ell\in\cE_k}\gamma_\ell]$,
which leads to the cover presented in
Theorem~\ref{thm:dspc}.

\section{The Bloch sphere and Reference POVM}
\label{app:sic}

The Pauli matrices are orthogonal for the trace inner product:
$\tr(H_{i}H_{i^\prime})=0$ if $i \neq i^\prime$, and $2$ otherwise. As
such, $\Phi_k= \sum_{i\in\{0\}\cup[3]}\phi_{k,i}H_i$ with
$\phi_{k,i}:=\frac{1}{2}\tr(\Phi_k H_i)$, and
$x^*\Phi_k x = \textstyle\sum_{i\in\{0\}\cup[3]} \phi_{k,i}
\nu_i(x)=T_k(v(x))$, where $\nu_i(x):=x^*H_i x$. The range of
$v(x)=(\nu_1(x),\nu_2(x),\nu_3(x))\in\R^3$ over unit $x\in\C^2$ is the
Bloch sphere~\cite[Sec.~1.2]{NC10}. The image of this under $T_k$ is
thus $\num(\Phi_k)$.

Given~\eqref{eq:referenceU}, direct calculation yields
$\tilde{v}_1=v(\tilde{u}_1)=(0,0,1)$, and
$\tilde{v}_m=v(\tilde{u}_m)=\big(\tfrac{2\sqrt{2}}{3}\cos\vartheta_m,\
\tfrac{2\sqrt{2}}{3}\sin\vartheta_m,\ -\tfrac{1}{3}\big)$ with
$\vartheta_m:=\tfrac{2\pi(m-2)}{3}$ for $m\in\{2,3,4\}$. These four
Bloch vectors satisfy $\tilde{v}_{m}^*\tilde{v}_{m^\prime}=-\frac{1}{3}$ for
$m\neq m^\prime$, \begin{align} \label{eq:refBlochVecSum} \textstyle
  \sum_{m\in[4]}\tilde{v}_m=(0,0,0) ~\text{ and }~
  \sum_{m\in[4]}\tilde{v}_m \tilde{v}_m^*=\frac{4}{3}I_3.
\end{align} 
The vectors $3\tilde{v}_m$ lie on the sphere of radius $3$, at equal
distances from each other, whereby $\Co(\{3\tilde{v}_m:m\in[4]\})$
is a regular tetrahedron~\cite{Renes04}.  The face opposite
$3\tilde{v}_m$ touches the Bloch sphere at $-\tilde{v}_m$, making the
encapsulation tight.

For $m\in[4]$,
$\tilde{u}_m \tilde{u}_m^* = \sum_{i\in\{0\}\cup[3]} \frac{1}{2}\!
\tr(H_i \tilde{u}_m\tilde{u}_m^*) H_i= \frac{1}{2}
\sum_{i\in\{0\}\cup[3]}\! \tr(\tilde{u}_m^* H_i \tilde{u}_m) H_i =
\frac{1}{2} \sum_{i\in\{0\}\cup[3]} \!\nu_i(\tilde{u}_m) H_i$. As
such,
$\textstyle\sum_{m\in[4]}\frac{1}{2} \tilde{u}_m
\tilde{u}_m^*=\frac{1}{4}\sum_{m\in[4]}\sum_{i\in\{0\}\cup[3]}\!
\nu_i(\tilde{u}_m)H_i
=H_0+\frac{1}{4}\sum_{i\in[3]}(\sum_{m\in[4]}\nu_i(\tilde{u}_m))H_i=I_2$,
since $\nu_0(\tilde{u}_m)=1$, $H_0=I_2$, and
$\sum_{m\in[4]}\nu_i(\tilde{u}_m)=0$ for $i\in[3]$ by the first part
of~\eqref{eq:refBlochVecSum}. As per~\eqref{eq:complete},
$\{\frac{1}{2}\tilde{u}_m\tilde{u}_m^*:m\in[4]\}$ is therefore a POVM,
\mc{which is also minimally informationally complete by linear
independence.} Further, with $\tilde{\varphi}_{k,m}=T_k(3\tilde{v}_m)$,
it follows from~\eqref{eq:refBlochVecSum} that
$\sum_{m\in[4]}\tilde{\varphi}_{k,m} = \sum_{m\in[4]} (\phi_{k,0} +
3\tilde{v}_m^* (\phi_{k,1},\phi_{k,2},\phi_{k,3})) = 4\phi_{k,0}$ and
$\sum_{m\in[4]}\tilde{\varphi}_{k,m}
\tilde{v}_{m}=\phi_{k,0}\sum_{m\in[4]}\tilde{v}_m + 3(\sum_{m\in[4]}
\tilde{v}_m \tilde{v}_m^*)(\phi_{k,1},\phi_{k,2},\phi_{k,3}) =
4(\phi_{k,1},\phi_{k,2},\phi_{k,3})$. Hence,
$\sum_{m\in[4]}\tilde{\varphi}_{k,m}\,\frac{1}{2} \tilde{u}_m
\tilde{u}_m^*= \frac{1}{4}(\sum_{m\in[4]}\tilde{\varphi}_{k,m})H_0+
\textstyle\frac{1}{4}\sum_{i\in[3]}\big(\sum_{m\in[4]}\tilde{\varphi}_{k,m}\nu_i(\tilde{u}_m)\big)
H_i= \sum_{i\in\{0\}\cup[3]}\phi_{k,i}H_i=\Phi_k$, which makes
$(\tilde{\varphi}_{k,m})_{m\in[4]}$ a corresponding representation of $\Phi_k$.

\section{Proof of Theorem~\ref{thm:aligned}}
\label{app:aligned}
Unitary $V_{k,\theta}$ in~\eqref{eq:Vrot} rotates
$v_{k,1}(\theta)=v(\mc{u_{k,1}(\theta)})$ to align with
$b_{k,\theta}\in\mc{\R^3}$, the direction in which
$\re(\exp(j\theta)T_k(\cdot))$ is maximized. The other three vertices
then lie on \mc{a plane tangent to} the Bloch sphere where
$\re(\exp(j\theta)T_k(\cdot))$ is minimized.  If $b_{k,\theta}=0$,
then $\re(\exp(j\theta)T_k(w))=\re(\exp(j\theta)\phi_{k,0})$ for all
$w\in\R^3$, and all terms in Theorem~\ref{thm:aligned} are
equal. Suppose $b_{k,\theta}\neq0$.

The rotated vectors $u_{k,m}(\theta)=V_{k,\theta}\tilde{u}_m$ make a
POVM because $V_{k,\theta}$ is unitary:
$\sum_{m\in[4]}\mu_{k,m}
u_{k,m}(\theta)u_{k,m}(\theta)^*=\sum_{m\in[4]} \mu_{k,m}
V_{k,\theta}\tilde{u}_m\tilde{u}_m^*V_{k,\theta}^*=V_{k,\theta}V_{k,\theta}^*
= I_2$. Moreover, the corresponding
$v_{k,m}=v(u_{k,m}(\theta))=R_{k,\theta} \tilde{v}_m$ for an
orthogonal matrix rotation
$R_{k,\theta}\in\R^{3\times 3}$~\cite[Sec.~4.2]{NC10}. As such, the
identities in~\eqref{eq:refBlochVecSum} hold with $v_{k,m}(\theta)$ in
place of $\tilde{v}_m$, and $(\varphi_{k,m})_{m\in[4]}$ with
$\varphi_{k,m}=T_k(\mc{3v_{k,m}(\theta)})$ is a \mc{correspondingly}
turned representation of $\Phi_k$.

If $\beta_{k,3}=-1$, then
$u_{k,1}(\theta)=V_{k,\theta}\tilde{u}_1=H_1 \tilde{u}_1=(0,1)$, with
Bloch vector
$v_{k,1}(\theta)=(0,0,-1)=\|b_{k,\theta}\|^{-1}b_{k,\theta}$.  When
$\beta_{k,3}\neq-1$, direct calculation yields
$u_{k,1}(\theta)
=V_{k,\theta}\tilde{u}_1=\tfrac{1}{\sqrt{2(1+\beta_{k,3})}}(1+\beta_{k,3},\,
\beta_{k,1}+j\beta_{k,2})$, with Bloch vector
$v_{k,1}(\theta)=(\beta_{k,1},\beta_{k,2},\beta_{k,3})=\|b_{k,\theta}\|^{-1}b_{k,\theta}$. Finally,
for $w\in\R^3$,
\[
\re(\exp(j\theta)T_k(w))=\re(\exp(j\theta)\phi_{k,0})+b_{k,\theta}^*w.
\]
By the Cauchy--Schwarz inequality, the minimum over the Bloch sphere
is $\re(\exp(j\theta)\phi_{k,0})-\|b_{k,\theta}\|$, attained at
$w=-v_{k,1}(\theta)=-\|b_{k,\theta}\|^{-1}b_{k,\theta}$.  This minimum
value is $\min\re(\exp(j\theta)\num(\Phi_k))$, since the Bloch sphere
under $T_k$ is $\num(\Phi_k)$, as shown in Appendix~\ref{app:sic}. At
the vertices, $b_{k,\theta}^*(3v_{k,1}(\theta))=3\|b_{k,\theta}\|$,
and
$b_{k,\theta}^*(3v_{k,m}(\theta))=3\|b_{k,\theta}\|\,v_{k,1}(\theta)^*v_{k,m}(\theta)=-\|b_{k,\theta}\|$
for $m\in\{2,3,4\}$. So $\re(\exp(j\theta)\varphi_{k,m}(\theta))$
equals the minimum for $m\in\{2,3,4\}$, and
$\re(\exp(j\theta)\varphi_{k,1}(\theta))$ exceeds it by
$4\|b_{k,\theta}\|$.

\section*{\mc{Acknowledgement}}

AI tools were used to assist with visualization coding for the figures,
and technical validations. The ideas, mathematical abstractions, proofs, and
narrative were all developed by the authors directly without input
from any LLM.


\begin{thebibliography}{99}
\bibitem{Willems10} J.~C.~Willems, ``Ports and terminals,'' in
  \emph{Perspectives in Mathematical System Theory, Control, and
    Signal Processing} (Lect. Notes Control Inf.
  Sci., vol.~398),
  Springer, 2010, pp.~27--36.
\bibitem{Harnefors07} L.~Harnefors, ``Modeling of three-phase dynamic
  systems using complex transfer functions and transfer matrices,''
  \emph{IEEE Trans. Ind. Electron.}, vol.~54, no.~4, pp.~2239--2248,
  2007.
\bibitem{Gu21} Y.~Gu, Y.~Li, Y.~Zhu, and T.~C.~Green,
  ``Impedance-based whole-system modeling for a composite grid via
  embedding of frame dynamics,'' \emph{IEEE Trans. Power Syst.},
  vol.~36, no.~1, pp.~336--345, 2021.
\bibitem{Gardonio97} P.~Gardonio, S.J.~Elliott, R.J.~Pinnington,
  ``Active isolation of structural vibration on a
  multiple-degree-of-freedom system, Part I,'' \emph{J. Sound
    Vibrat.}, vol.~207, no.~1, pp.~61--93, 1997.
\bibitem{ZDG96} K.~Zhou, J.~C.~Doyle, and K.~Glover, \emph{Robust and
    Optimal Control}. Upper Saddle River, NJ, USA: Prentice Hall,
  1996.
\bibitem{MR97} A.~Megretski and A.~Rantzer, ``System analysis via
  integral quadratic constraints,'' \emph{IEEE Trans. Autom. Control},
  vol.~42, no.~6, pp.~819--830, 1997.
\bibitem{Vin01} G.~Vinnicombe, \emph{Uncertainty and Feedback:
    $\mathcal{H}_\infty$ Loop-Shaping and the $\nu$-Gap
    Metric}. London, U.K.: Imperial College Press, 2001.
\bibitem{LV06} I.~Lestas and G.~Vinnicombe, ``Scalable decentralized
  robust stability certificates for networks of interconnected
  heterogeneous dynamical systems,'' \emph{IEEE Trans. Autom.
    Control}, vol.~51, no.~10, pp.~1613--1625, 2006.
\bibitem{Lestas12} I.~Lestas, ``Large scale heterogeneous networks,
  the Davis--Wielandt shell, and graph separation,'' \emph{SIAM
    J. Control Optim.}, vol.~50, no.~4, pp.~1753--1774, 2012.
\bibitem{NC10} M.~A.~Nielsen and I.~L.~Chuang, \emph{Quantum
    Computation and Quantum Information}, Cambridge, U.K.: Cambridge
  Univ. Press, 2010.
\bibitem{Renes04} J.~M.~Renes, R.~Blume-Kohout, A.~J.~Scott, and
  C.~M.~Caves, ``Symmetric informationally complete quantum
  measurements,'' \emph{J. Math. Phys.}, vol.~45, no.~6,
  pp.~2171--2180, 2004.
\bibitem{JK10} U.~T.~J\"onsson and C.-Y.~Kao, ``A scalable robust
  stability criterion for systems with heterogeneous LTI components,''
  \emph{IEEE Trans. Autom. Control}, vol.~55, no.~10, pp.~2219--2234,
  2010.
\bibitem{KR14} S.~Z.~Khong and A.~Rantzer, ``Scalable stability
  conditions for heterogeneous networks via integral quadratic
  constraints,'' in \emph{Proc. Eur. Control Conf.}, Strasbourg,
  France, 2014, pp.~2863--2867.
\bibitem{PV17} R. Pates and G. Vinnicombe, ``Scalable design of
  heterogeneous networks,'' \emph{IEEE Trans. Autom. Control},
  vol.~62, no.~5, pp.~2318--2333, 2017.
\bibitem{Huang24} L.~Huang, D.~Wang, X.~Wang, H.~Xin, P.~Ju,
  K.~H.~Johansson, and F.~D\"orfler, ``Gain and phase: Decentralized
  stability conditions for power electronics-dominated power
  systems,'' \emph{IEEE Trans. Power Syst.}, vol.~39, no.~6,
  pp.~7240--7256, 2024.
\bibitem{HHHDL26} V.~H\"{a}berle, X.~He, L.~Huang, F.~Dörfler, and S.~Low, ``Decentralized parametric stability certificates for grid-forming converter control,'' \emph{IEEE Trans. Control Netw. Syst.}, vol.~13, no.~3, pp. 1271--1283, 2026.
\bibitem{WCGPM26} Z.~Wang, S.~Chatterjee, S.~Geng, R.~Pates, and E.~Mallada, ``Decentralized Stability Certificates in IBR-Dominated Grids: The Role of the Network State,'' in \emph{Proceedings of the PowerUp
    Conference}, 2026.
\bibitem{BAD25} E.~Baron-Prada, A.~Anta, and F.~D\"{o}rfler, ``On decentralized stability conditions using scaled relative graphs,'' \emph{IEEE Control Syst. Lett.}, vol.~9, pp.~691--696, 2025.
\bibitem{PV12} R.~Pates and G.~Vinnicombe, ``Stability certificates
  for networks of heterogeneous linear systems,'' in \emph{Proc. 51st
    IEEE Conf. Decision Control}, Maui, HI, USA, 2012, pp.~6915--6920.
\bibitem{HL26} L.~Hallinan and I.~Lestas, ``Decentralised
  plug-and-play stability conditions for AC grids---Part~I,''
  arXiv:2609.07315, 2026.
\bibitem{HornJohnson} R.~A.~Horn and C.~R.~Johnson, \emph{Matrix
    Analysis}, 2nd~ed. Cambridge, U.K.: Cambridge Univ. Press, 2013.
\bibitem{BV04} S.~Boyd and L.~Vandenberghe, \emph{Convex
    Optimization}. Cambridge, U.K.: Cambridge Univ. Press, 2004.
\bibitem{CZ95} R.~Curtain and H.~Zwart, \emph{An Introduction to
    Infinite-Dimensional Linear Systems Theory}, ser. Texts in Applied
  Mathematics. New York, NY, USA: Springer, 1995.
\bibitem{HornJohnsonTopics} R.~A.~Horn and C.~R.~Johnson, \emph{Topics
    in Matrix Analysis}. Cambridge, U.K.: Cambridge Univ. Press, 1991.
\bibitem{Schur11} I.~Schur, ``Bemerkungen zur Theorie der
  beschr\"ankten Bilinearformen mit unendlich vielen
  Ver\"anderlichen,'' \emph{J. Reine Angew. Math.}, vol.~140,
  pp.~1--28, 1911.
\end{thebibliography}
\end{document}